\documentclass[12pt]{iopart}

\usepackage[latin1]{inputenc}
\usepackage[dvips]{graphicx}
\begin{document}

\title{Good practices for a literature survey are not followed by authors while preparing scientific manuscripts}

\author{D. R. AMANCIO$^1$, M. G. V. NUNES$^2$, O. N. OLIVEIRA JR.$^1$ and L. da F. COSTA$^1$ } 
\address{$^1$  Institute of Physics of São Carlos \\  
	University of São Paulo, P. O. Box 369, Postal Code 13560-970 \\
	São Carlos, São Paulo, Brazil \\}

\address{$^2$ Institute of Mathematics and Computer Science \\
 	University of São Paulo, P. O. Box 668, Postal Code 13560-970 \\
	São Carlos, São Paulo, Brazil  }

\ead{diego.amancio@usp.br, gracan@icmc.usp.br, chu@ifsc.usp.br, luciano@ifsc.usp.br}

\begin{abstract}
The number of citations received by authors in scientific journals has become a major parameter to assess individual researchers and the journals themselves through the impact factor. A fair assessment therefore requires that the criteria for selecting references in a given manuscript should be unbiased with respect to the authors or the journals cited. In this paper, we advocate that authors should follow two mandatory principles to select papers (later reflected in the list of references) while studying the literature for preparing a manuscript: i) consider similarity of contents with the topics investigated, lest related work should be reproduced or ignored; ii) perform a systematic search over the network of citations including seminal or very related papers.  We use formalisms of complex networks for two datasets of papers from the arXiv repository to show that neither of these two criteria is fulfilled in practice. Representing the texts as complex networks we estimated a similarity index between pieces of texts and found that the list of references did not contain the most similar papers in the dataset. This was quantified by calculating a consistency index, whose maximum value is one if the references in a given paper are the most similar in the dataset. For the areas of ``complex networks'' and ``genetics'', the consistency index was only 0.19-0.29 and 0.30-0.47, respectively. To simulate a systematic search in the citation network, we employed a random walk search (i.e.\ diffusion). The frequency of visits to the nodes (papers) in the network had a very small correlation with either the actual list of references in the papers (Pearson coefficient = 0.075) or with the number of downloads (Pearson = 0.165) from the arXiv repository. Therefore, apparently the authors and users of the repository did not perform a systematic search over the network of citations. Based on these results, we propose an approach that we believe is fairer for evaluating and complementing citations of a given author, effectively leading to a virtual
scientometry. 

\end{abstract}

\maketitle

\section{Introduction}
The advance of knowledge is founded and critically dependent on the broad dissemination of novel approaches and results, which allows other scientists and practitioners to analyze reported results to validate and complement their investigations. The primary objective of any scientific publication is therefore to be \emph{read}, \emph{tried}, and \emph{cited} by as many people as possible. Indeed, articles have been evaluated in terms of the citations they motivate, while journals are typically rated according to the impact factors reflecting the number of citations to their articles. Citations have been a major factor since the 1920s~\cite{Gross}, and subjects such as this are now analyzed in scientometrics, which studies the relationship between areas of knowledge and the evolution of science~\cite{katy}. Of course, the success of a paper in being read and cited varies enormously owing to several factors, including the renown of the journal and the eminence of the authors and their institutions. Strictly speaking, such a success should depend not only on the quality, originality, completeness and clarity of a specific paper, but moreover on the degree of relationship and overlap with the investigation being reported. For all the papers that are strongly similar or related to a current investigation should be read, and potentially cited. However, with the limited time available to any researcher for seeking and reading, the related works have to be somehow filtered by using some limiting criteria. Though unavoidable, this implies that potentially important publications are overlooked~\cite{malcom,lilien}, which may undermine the efficiency of the whole system, in the sense that painstaken, costly efforts are repeated or ignored.

We take the view that little attention has been given to the procedures of selecting publications for guiding the research and preparing a list of references~\cite{abuses}. In this paper we suggest two criteria for such a selection. The first is that similar, strongly related works should be selected, and the second is that the authors should do a systematic search over related publications and their citations. We check whether these criteria are fulfilled by using complex networks~\cite{barabasi,stat,newm} and natural language processing formalisms (for the use of complex networks in natural language processing, see~\cite{cancho1,cancho2,sigman,whatname}). Two datasets containing 700 articles each from the arXiv\footnote{http://arXiv.org} repository, for the areas of complex networks and genetics, were used to obtain two networks for each area: (i) the traditional citation networks, where each article is a node and citations become directed edges between them; and (ii) a network obtained by the overlap between the contents of pairs of articles. These networks are henceforth referred to as \emph{citation} and \emph{overlap}, being directed and undirected, respectively. Concerning the overlap network, each article was modeled as a complex network in order to extract the relations of similarity. The model used (see methodology), which basically connects adjacent words after a pre-processing step, was chosen because of its success in other studies on Natural Language Processing, such as automatic text assessment~\cite{quality}, automatic summarization strategies~\cite{summ} and automatic machine translation assessment~\cite{tam}. After defining these two networks one may quantify the number of: (a) articles which are related and cited; (b) articles that are related but not cited; and (c) articles that are loosely related but are cited nonetheless. We shall show that the analysis of these numbers indicates that the similarity criterion for selecting references is not obeyed. We also perform a random walk through the citation networks to simulate a systematic search by an author, whose results are used to infer that the second criterion is not obeyed either. In addition to discussing the possible causes and implications of these results, we suggest a virtual citation approach for complementing the relationships between articles, which gives rise to a virtual scientometry. 

\section{Methodology}

In our experiments, the relationships (similarity and citation) between two articles are modeled as complex networks. A network is defined as a data structure comprising a set of nodes linked by edges. The set of edges and nodes can be represented as a matrix W$_{ij}$, where the presence of an edge between two nodes \textit{i} and \textit{j} with weight \textit{p} infers W$_{ij}$ = \textit{p} and the absence of an edge implies W$_{ij}$ = 0. If there is no order distinction to link two nodes (\textit{i} $\rightarrow$ \textit{j} is the same as \textit{j} $\rightarrow$ \textit{i}), then W$_{ij}$ = W$_{ji}$ is always true. If two nodes are connected by an edge, they are said to be adjacent. If two edges are associated with the same node, they are called adjacent edges. A sequence of adjacent edges defines a walk over the network. The length of a walk is defined as the number of the edges of the walk.
The networks were built using a corpus comprising 700 articles about complex networks (or scale free networks) and 700 articles about genetics from the arXiv publications base. The articles were randomly selected from the database, which contained the keywords ``complex networks" or ``scale-free networks" and ``genetics" in the title or in the abstract. Only the most current version of the manuscripts was considered. It should be mentioned that any search on the arXiv site returns a set of articles that are not necessarily in chronological order of publication.

\begin{figure*}
\begin{center}
\fbox{\includegraphics[width=0.45\textwidth]{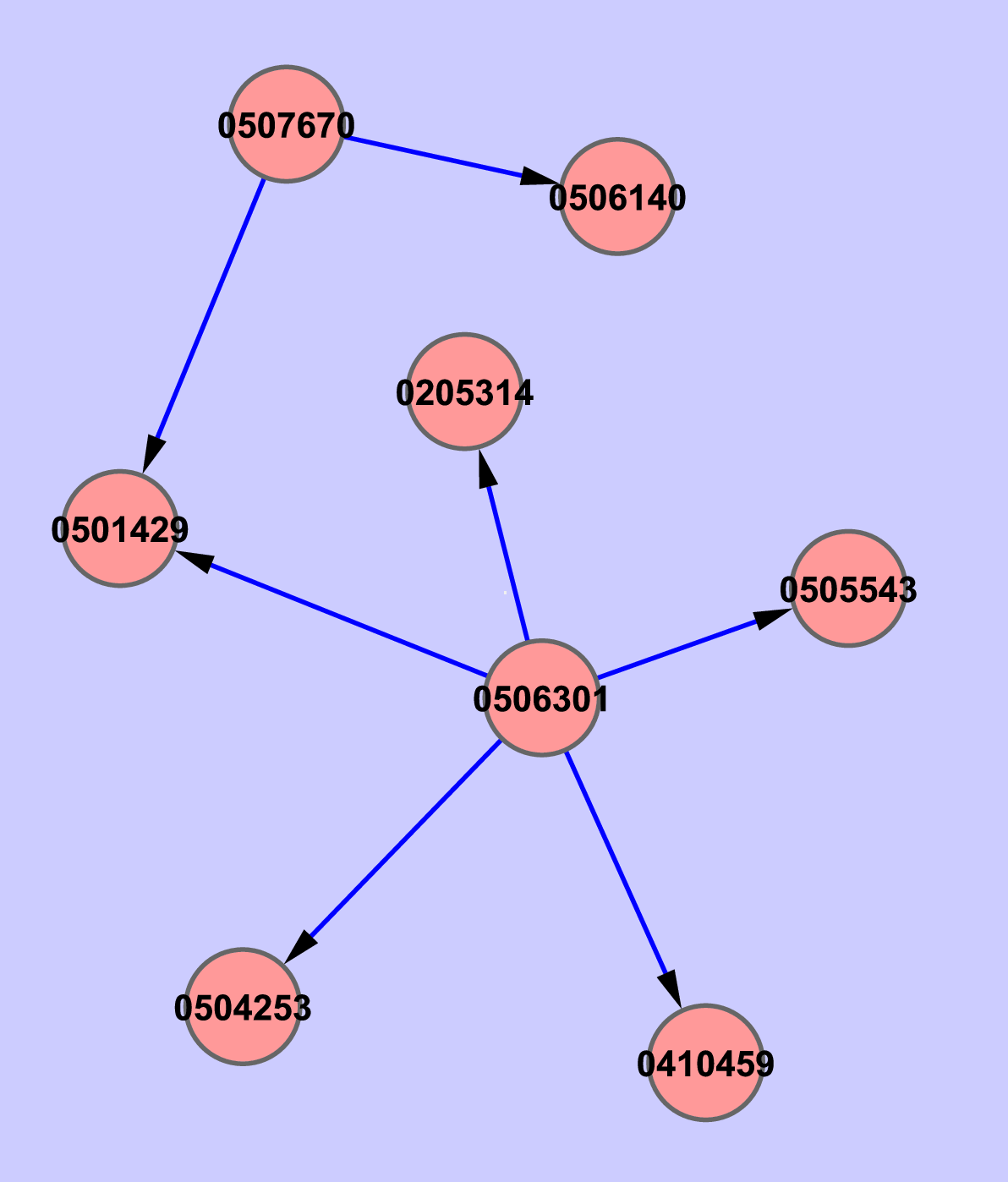}}
\end{center}
\caption{Visualization of a citation subnetwork for the complex network subject. While article cond-mat/0506301 cites five articles within the arXiv base, the article cond-mat/0507670 cites only two articles within the same base.}
\label{fig.1}
\end{figure*}

The citation network W$^{cit}_{ij}$, which can be considered as a modified social network (the relationship between people is replaced by the relationship between articles)~\cite{doreian} and is known to follow a power law distribution~\cite{cita}, stores all the information about citations among articles, where each article is a vertex and each edge represents a citation.  If article $i$ cites article $j$ then there will be a directed edge network represented as $i$ $\rightarrow$ $j$. Figure~\ref{fig.1} illustrates some articles and their references.

\begin{figure*}
\begin{center}
\fbox{\includegraphics[width=0.45\textwidth]{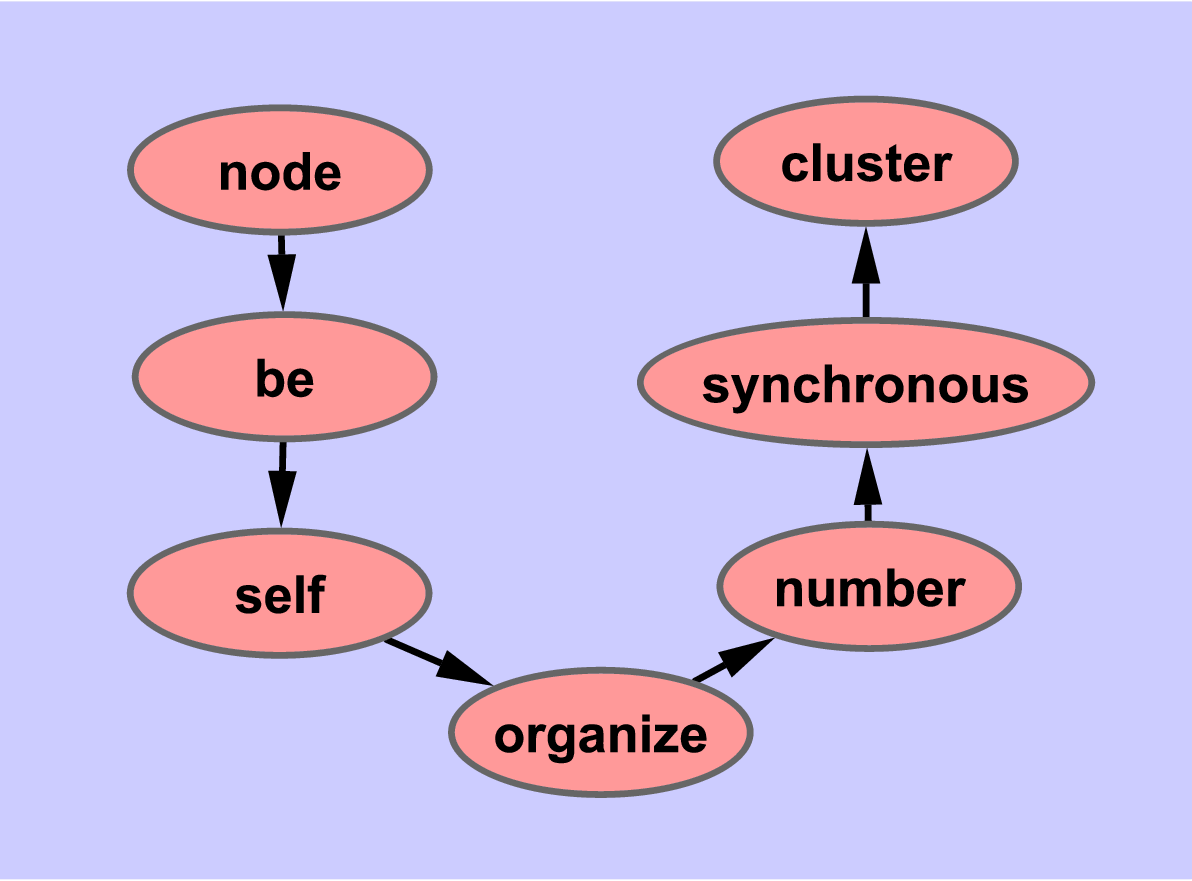}}
\end{center}
\caption{Network obtained from the sentence ``Nodes are self organized into a number of synchronous clusters".}
\label{fig.2}
\end{figure*}

For building the overlap network, each file representing a paper in the database was pre-processed in order to remove the tags from \LaTeX\ markup language, so that specific terms from LaTex do not affect the calculation of the similarity index (see below). The overlap network W$^{sim}_{ij}$ is undirected and comprises vertices representing the articles (as in a network of citations) and edges that connect two articles with some degree of similarity. Specifically, for each edge, a weight proportional to the similarity between the two vertices in this relation is defined. To avoid a highly connected network, we used a threshold so that all weights below the threshold were eliminated. For obtaining the similarity between two papers (vertices), a two-stage procedure was adopted: (i) modeling each text as a complex network~\cite{lantiqnetw}; and (ii) comparing the corresponding networks. In modeling the text as a network, the stopwords were removed. Moreover, the remaining words were lemmatized to combine concepts with the same canonical form, but with different inflections. Additionally, the text was labeled using the MXPost part-of-speech Tagger~\cite{aires} based on the Ratnaparki's model~\cite{ratim}, which helps to resolve problems of ambiguity. This is useful because the words with the same canonical form and the same meaning are grouped into a single node, while words that have the same canonical form but distinct meanings generate distinct nodes. This pre-processing is done by accessing a computational lexicon~\cite{nunes}, where each word has an associated rule for the generation of the canonical form. 
The structure that represents the network derived from a text is a weighted adjacency matrix. After the pre-processing, the N words represented the network nodes and the resulting sequence of words were used to create the edges: for each pair of consecutive words there was a corresponding edge in the network. The edges also had weights, which indicated the number of times that the associations of words appeared in the text. This network was stored as a directed adjacency matrix, named W. The latter was initialized with zero elements, and as each word pair was read from the text, $W_{ij}$ was incremented. In other words, the corresponding edge had its weight increased. Figure~\ref{fig.2} shows the network obtained from the sentence: ``Nodes are self organized into a number of synchronous clusters". 

\subsection{Topology-based Similarity Index (TSI) and Consistency Index (CI) }
It is well accepted that two pieces of text may be considered as similar if they share a large number of common concepts (represented by words), also taking into account the co-occurrence of concepts. In order to quantify the extent of similarity we propose an index that is calculated by comparing the networks of the texts considered, in which the metrics selected are believed to capture the network topology. Therefore, the index involves not only the common concepts (nodes) of the two networks, but also the vicinity of such nodes. Formally, the similarity index between two texts was obtained as follows. Let \textit{s} be the number of occurrences of a word \textit{p} in the first pre-processed text and \textit{t} the number of occurrences of the same word in the second pre-processed text. The factor $\alpha_p$ defined in \ref{eq.1} is a first measure of similarity.

\begin{equation}
			\alpha_p  = \min (s,t) 
			\label{eq.1}
\end{equation}

where min(s,t) stands for the minimum between $s$ and $t$. 

We then use information from the adjacency word network. Let $v_s$ be the vertex of the adjacency network representing the word \textit{p} in the first pre-processed text and $v_t$ the vertex of the adjacency network representing the same word in the second pre-processed text. If the vertex $v_s$ has $\kappa$ neighbors in common with the neighbors of $v_t$ and if $W_s$ represents the first network and $W_t$ the second, the $\beta_p$ factor is defined as in \ref{eq.2}, so that the number of common neighbors is captured.

\begin{equation}
			\beta_p  = \frac{\sum_{v=1}^{\kappa} \min \Big{\{}W_s(v_s,v), W_t(v_t,v)\Big{\}} }{\sum_{v=1}^{\kappa} W_s(v_s,v) + \sum_{v=1}^{\kappa} W_t(v_t,v) } 
		\label{eq.2}
\end{equation}

With these definitions, the similarity index for the \textit{p} word, referred to as $I_p$, is given in \ref{eq.3}.
\begin{equation}
	I_p  = \alpha_p \cdot \beta_p 
	\label{eq.3}
\end{equation}

The global similarity index \textit{TSI} between these two texts is then defined in \ref{eq.4}, where $N_s$ is the number of word occurrences in the first pre-processed text and $N_t$ is the same for
the second pre-processed text.
		\begin{equation}
			TSI  = \frac{1}{N_s + N_t} \sum{}{} I_p
			\label{eq.4}
		\end{equation}

\begin{figure*}[ht]
\centerline{\fbox{\includegraphics[width=0.50\textwidth]{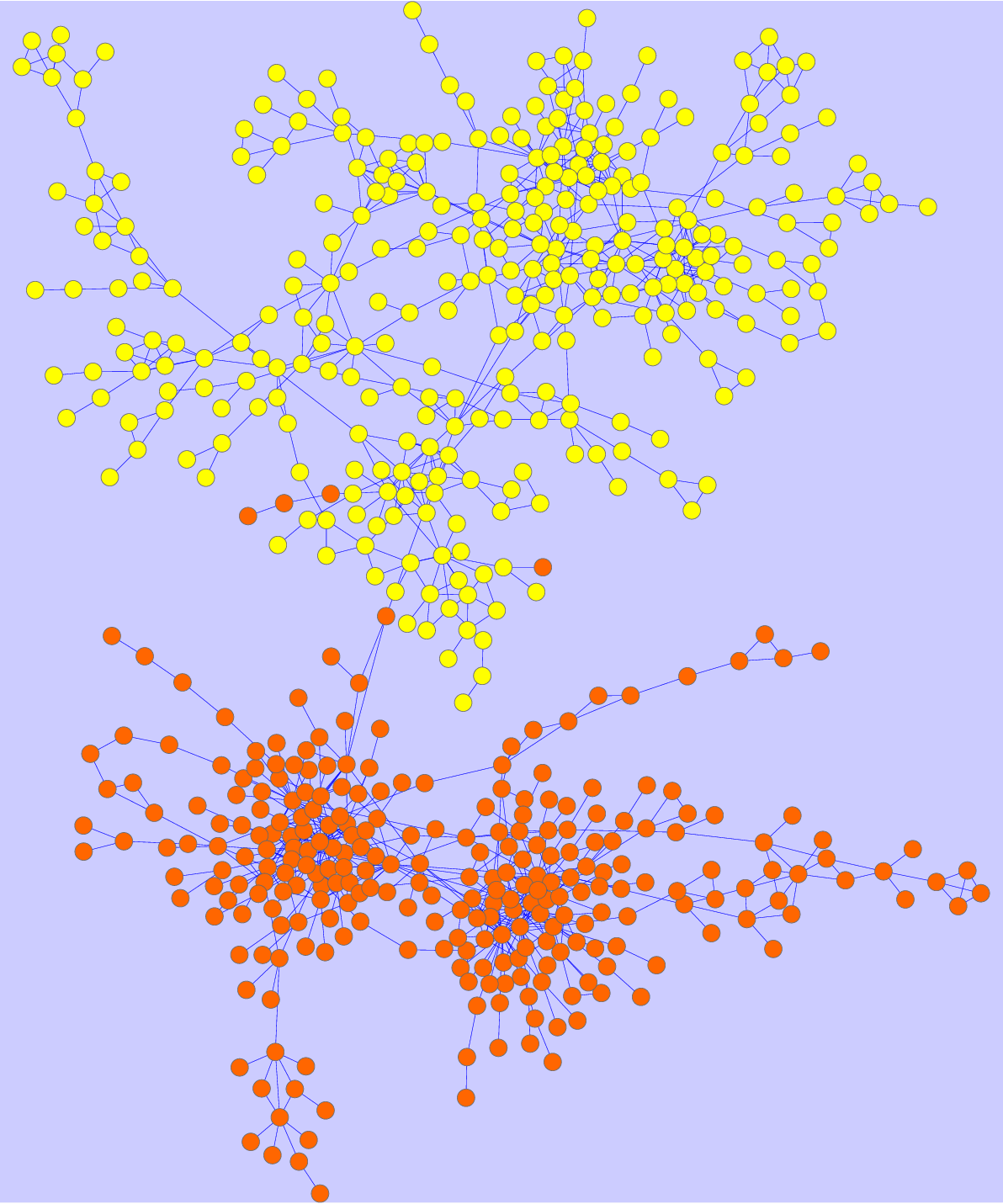}}}
\caption{Visualization of a similarity network (largest component) for the complex network (orange) and genetics (yellow) subject. The few connections between vertices of different colors (different subjects) show that the similarity index connects similar articles.}
\label{fig.3}
\end{figure*}

Assessing similarity between texts is highly subjective. Nevertheless, we believe that the index defined above is at least reasonable, as indicated in Figure \ref{fig.3}, where a visualization using the software Cytoscape\footnote{www.cytoscape.org} is provided of the networks for genetics and complex networks. When linked as a single network with the edges between vertices being dependent on the similarity index, one notes that the vertices of the same color remain practically separated from the other color. The color represents a given subject, and little overlap is seen in texts of different subjects, thus pointing to an effective separation via the similarity index.

Because in this paper we advocate that the citations in a given paper should be based on the similarity of contents with the manuscript being written, we quantified whether this expectation is fulfilled by using a consistency index. By consistent we mean that while selecting papers to cite, the authors considered the similarity of content. The consistency index $CI$ was conceived to vary from 0 to 1, where 1 corresponds to the case where all the references cited in the papers of the database are contained in the selected database (see the definition below). The value 0 is for the case where the citations in the papers do not belong to the selected database. We therefore refer to the total database as that containing the $\eta$ papers from arXiv used in the citation network (we used $\eta$ = 700), while the selected database comprises the papers cited which belong to the total database.
The procedure to obtain $CI$ is as follows. For each paper \textit{i} in the database, one obtains the similarity threshold for which the number of similar papers in the database is equal to the number of references in paper \textit{i}. The papers from the arXiv database satisfying this requirement for all the papers from \textit{i}=1 to \textit{i}=$\eta$ comprise the so-called selected database. To illustrate: suppose paper i has 20 references. This paper will contribute with 20 papers - the most similar to paper i - to the selected database. Let $x_i$ be the number of references of paper \textit{i} which belong to the selected database. Let $y_i$ be the number of references of paper \textit{i} which belong to the complete arXiv database. The ratio $x_i$/$y_i$ therefore gives the fraction of papers cited that are actually similar to the original one. For instance, suppose that 8 out of the 20 references in paper i belong to the arXiv database (complete database with $\eta$ papers). Then, $y_i$ = 8. If 4 of these 8 references appear in the selected database (the most similar given the threshold mentioned above), then $x_i$ = 4, and the ratio is 0.5. The Consistency Index will be the sum of $x_i$/$y_i$, from i = 1 to $\eta$, divided by $\eta$, as shown in ~\ref{eq.5}. High values will mean that the references are indeed similar to the original paper. 

\begin{equation}
	CI = \frac{1}{\eta} \sum_{i=1}^{\eta} \frac{x_i}{y_i}
	\label{eq.5}
\end{equation}

Now, it is possible that $CI$ = 1 and the similarity is not high. Let us exemplify with an extreme case. Suppose a paper with $x_i$ = 3 and $y_i$ = 3, but the 3 references are the least similar in the selected database (just above the threshold). In order to account for that, we created another parameter $CI'$, which considers the order in which the papers appear (i.e. 1 for the most similar and 0 for the least similar - the last included in the selected database).

\subsection{Simulating a systematic search over the citation network}
In order to verify the theoretical access frequency to the articles, the dynamics of the citation networks was analyzed using the traditional random walk mechanism, where the next edge is chosen with uniform probability among the adjacent edges. Then, the dynamics of the matrix W$_{ij}$ can be represented as a Markov Chain~\cite{markov}. At each step the network may change its state from the current to another state (or remain in the same state) according to the probability distribution assigned previously.  Here, we assume that the transition probabilities are associated with the probability of a reader to follow an edge, which is taken as proportional to the weight of such an edge. Thus, articles with high probability of access at the steady state (when time tends to infinity) are more likely to be visited in a random walk. Our hypothesis is that if authors perform a systematic search over the citation network, then the articles with the highest access frequency in the steady state are more likely to be consulted (and hence cited). To verify whether this criterion in selecting the papers was obeyed, we calculated the Pearson coefficient~\cite{pearson} between the actual number of citations (or downloads) and the expected access frequency from the random walk. 

\section{Results and Discussion}

The main thrust of this paper is to provide formalisms to verify whether possible criteria in selecting references for a paper being prepared are followed. We suggest that two mandatory principles to be followed by authors in a literature survey and in selecting publications to study, and eventually include in the list of references, are: i) choose strongly related papers, which could be done by the similarity of the contents; ii) perform a systematic search on the citation network involving seminal and strongly related publications. It is true that other criteria could also be used. For instance, the authors could seek the publications considered the most relevant in the topic under study, which could be identified by the number of citations or the reputation of the authors or the journal in which the papers appeared. But using this criterion would inevitably introduce a bias, which we wish to avoid. Besides, such relevant papers are likely to be found in the systematic search. Another possibility is that the authors may need to refer to papers that are only weakly related, but which may for instance provide a methodology from another field used in the research. While this would be certainly justifiable in the selection of references, it is probably of little relevance in terms of the statistics of citation networks, for only a small number of references would be generated if the methodology were not strongly related to the work being performed. Therefore, for the purposes of this paper, we need not consider the latter criterion.
In the following we present results obtained with the overlap and citation networks to show that the two criteria proposed here are not fulfilled.

\subsection{Consistency Index Analysis}
In order to check whether authors select similar papers to include in their list of references, we defined a consistency index $CI$ which would be 1 if all the references in a paper are the most similar in the dataset and zero when none of the references are among the most similar. The consistency indexes $CI$ and $CI'$ were computed for the corpus related to the subject of ``complex networks" and ``genetics", using either the full text or extracts from the summary and the introduction. The reason why these extracts were used was that the Introduction and Abstract could be the sections most representative of the topics in a paper. More specifically, by concentrating on these sections one could avoid the problem of weakly related works being included owing to the mere use of a methodology from another field of research, as mentioned above. The texts were represented by complex networks~\cite{lantiqnetw}, whose topology was used to obtain the similarity of contents between two pieces of text (see the Methodology). The results shown in tables~\ref{tab.3} and ~\ref{tab.3b} point to relatively low consistency indexes for both subjects, regardless of whether the full text or extracts were employed in computing the index. Indeed, in average only 1 out of 5 references appearing in an article on complex networks was really similar. As for the genetics subject, 1 out of 3 was similar. Interestingly, the value of $CI'$ is approximately half $CI$, which means that the citations were distributed regularly in terms of similarity. That is, they are neither concentrated among the most similar nor among the least similar. 
The conclusions above were not affected by using other methods for calculating the similarity index, such as ignoring the vertex neighborhood, using the full article or using only articles with many references within the base. In fact, ignoring $\beta_p$ (see the Methodology) corresponds to calculating the similarity based only on the number of shared words between two texts, without taking into account the topology of the network. These results confirm the hypothesis that other factors are more relevant than the similarity (e.g, it is known that the publication date is relevant, because older articles tend to be forgotten~\cite{cita,citb}).

\begin{table}
\caption{\label{tab.3}Consistency index computed for complex networks}
\begin{indented}
\item[]\begin{tabular}{@{}lll}
\br
\textbf{Description} 	& 	\textbf{$CI$} & \textbf{$CI'$} \\
\mr
Similarity index calculated with 	&  &	\\
complex networks metrics	for	&	0.20&	0.10\\
the introduction and abstract		&	&  \\
\mr
Similarity index calculated 	&  &	\\
with complex networks metrics	&	0.29	& 0.14	\\
for full articles					&	& \\
\mr
Similarity index calculated 	&  &	\\
by co-occurrence of words for 	&	0.19	& 0.10	\\
the introduction and abstract	&	& \\
\br
\end{tabular}
\end{indented}
\end{table}

\begin{table}
\caption{\label{tab.3b}Consistency index computed for genetics}
\begin{indented}
\item[]\begin{tabular}{@{}lll}
\br
\textbf{Description} 	& 	\textbf{$CI$} & \textbf{$CI'$} \\
\mr
Similarity index calculated with 	&  &	\\
complex networks metrics	for	&	0.34&	0.20\\
the introduction and abstract		&	&  \\
\mr
Similarity index calculated 	&  &	\\
with complex networks metrics	&	0.47	& 0.31	\\
for full articles					&	& \\
\mr
Similarity index calculated 	&  &	\\
by co-occurrence of words for 	&	0.33	& 0.21	\\
the introduction and abstract	&	& \\
\br
\end{tabular}
\end{indented}
\end{table}

\subsection{Systematic search over the citation network}
We have advocated in this paper that authors should perform a systematic search on the citation network involving seminal or related papers to the research being conducted. Here we mimic such search via a random walk search in the citation networks for the two subjects analyzed, namely ``complex networks" and ``genetics". In order to determine the relative frequency of access for each node in the citation network, the transition matrix $\tau$ is created by first redefining the citation matrix W$^{cit}_{ij}$ (selecting its greater component) as A$^{cit}_{ij}$ to reflect a reader's browsing on the network. The element A$^{cit}_{ij}$ has the value 1 if there is an edge \textit{i} $\rightarrow$ \textit{j} in the citation network W$^{cit}_{ij}$. If the edge \textit{i} $\rightarrow$ \textit{j} exists in W$^{cit}_{ij}$ but the edge \textit{j} $\rightarrow$ \textit{i} does not exist, then one considers this association in A$^{cit}_{ij}$ with weight 0.2, since it is easier for a reader to follow a reference in an article than to find out who cited this article. Hence, we assigned the lower weight to the less probable direction. The matrix $\tau$ is a stochastic matrix, derived from A$^{cit}_{ij}$ as in~\ref{eq.6}:

\begin{equation}
			\tau_{ij}  = \frac{{A}_{ij}^{cit}}{\sum_{j}^{}{A}_{ij} } 
			\label{eq.6}
\end{equation}

This matrix measures the probability of a walker visiting a node \textit{j} after being at node \textit{i}. We are interested in the stationary, or steady-state, distribution of probabilities of being at each node, i.e., one wants to know the probability of being at a given node after an infinite number of steps. This distribution, denoted as $\pi$, can be obtained by solving the equation $\tau$$\cdot$$\pi$ = $\pi$. In particular, $\pi$ is the eigenvector associated with the eigenvalue 1 of $\tau$, where the sum of all elements of $\pi$ is equal to 1 to reflect the sum of probabilities. It is known that for this distribution to be unique, $\tau$ needs to be irreducible, i.e. the network must be strongly connected, which happens when every node can reach every other node in the network through a finite path. Here this property was guaranteed as we selected the greater component of W$^{cit}_{ij}$.

\begin{figure*}
\begin{center}
\includegraphics[width=0.75\textwidth]{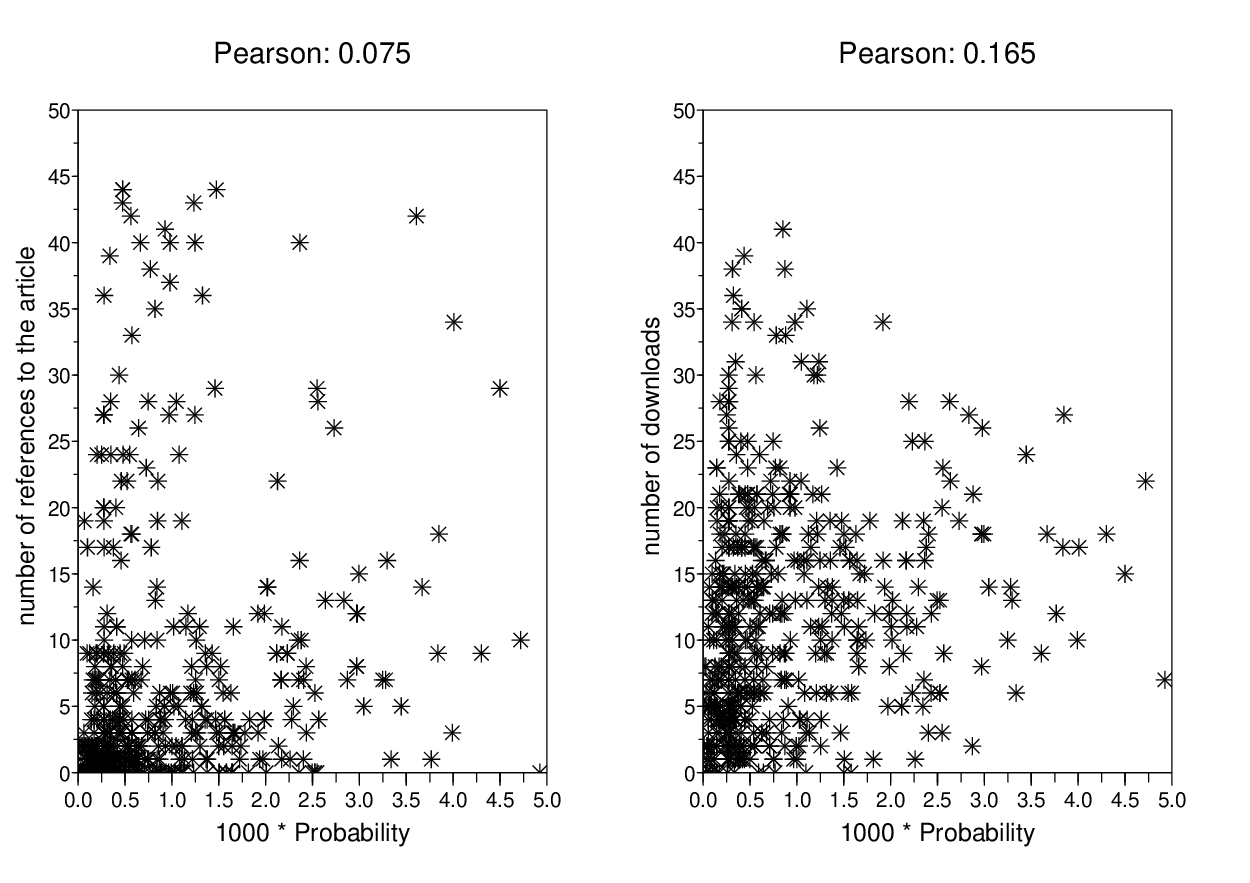}
\end{center}
\caption{Scatter plots confirm a weak correlation between the spectral measure computed from the citation network and the real access frequency.}
\label{fig.4}
\end{figure*}

The Pearson coefficient~\cite{pearson} was calculated to verify the correlation between the expected frequency of visits to the nodes (articles) and the actual citations and downloads for the papers in the arXiv repository. The latter information was obtained from the Citebase's site\footnote{http://www.citebase.org}, with the number of downloads and number of references to a given article. The results are illustrated in table~\ref{tab.4} and in Figure\ref{fig.4}.

\begin{table}
\caption{\label{tab.4}Correlation between the expected frequency of visits computed from the citation network and the real access frequency.}
\begin{indented}
\item[]\begin{tabular}{@{}ll}
\br
\textbf{Measure} 	& 	\textbf{Correlation} \\
\mr
Number of references to the article 			&  0.075	\\
Number of downloads									&	0.165	\\
\br
\end{tabular}
\end{indented}
\end{table}

The weak correlation means that researchers did not consult the references for selecting the papers to cite or to download; in other words, the criterion of a systematic search over the citation network was not fulfilled either. 

\subsection{Towards Virtual Scientometry}

The results obtained in our analysis clearly indicate, at least in the case of the arXiv database, that two of the most important criteria which should guide citations are not followed in practice. This is worrying because a basic premise of the scientific method, namely that related work should be analyzed to avoid reproduction and warrant originality, is not fulfilled. There is also the side effect that the citation statistics for authors and journals is likely to be strongly biased. Perhaps one should not be surprised with these conclusions. For in a survey sent to psychology journal editors, Croner~\cite{cronin} showed that more than 80\% of participants believed that authors of papers failed to cite relevant articles, since some citations seemed to be forged to get the attention of editors~\cite{vinkler}.  Additionally, there is also the problem of publication date, since there is a tendency to remember recent papers in detriment to older ones~\cite{wang, webera}. So, somewhat, publications are made not only taking into account the intellectual influences from scientific peers, but also non-scientific factors~\cite{bornman}. We believe that means should be devised to complement and/or correct the citation procedures normally adopted.  While it would be very difficult to change the established citation habits, the continuing advances in informatics and Internet now allow for innovative approaches to be implemented.  Here, we outline a computational approach which could be used to assist authors in identifying relevant references. Furthermore, it could be used in assessing impact of researchers and papers with the concept of virtual citations, to be defined below.

The automated tool for helping authors to identify relevant references could be a simple implementation of the methods reported in the previous sections. More specifically, a manuscript in preparation could be transformed into a network and its content would be compared to all documents in a given database of articles (e.g.\ arXiv) using the similarity indexes defined in this work. The output would be a list of related works ranked by decreasing order of overlap/similarity. The author may then check this list and identify potentially relevant works which could otherwise be overlooked. This software tool could also be used to obtain virtual citations for a paper already published, as follows. Given a research area for which a set of papers could be selected, as in the database from the arXiv repository used here, the number of items to be included could be fixed or depend on the similarity of contents among the papers. The virtual citations for a paper would be chosen above a threshold of similarity, which should be defined as to yield an average of citations that is equal to the actual number of references when considering all the papers in the database. Therefore, a paper that is similar to a large number of other papers in the database would receive many citations, as it supposedly deals with a hot topic. Using the citation network obtained from the database, as discussed in the subsection above, another possibility to evaluate the impact of a paper would be to calculate the frequency of visits in a random walk through the network. 

The two metrics obtained with the virtual citations and the random walk could be used to compare the impact of an author (or a specific paper) with other authors (or papers), with the following advantages. a) No bias existed in selecting the citations.  b) The comparison is straightforward with counterparts in the same field. Therefore, there is no such effect as a higher impact in a field that has a higher number of citations per paper than in other fields. 
It should be emphasized that we are by no means suggesting that the traditional system of citations should be replaced by the virtual citations and the metrics arising from the frequency of visits in the citation network, as proposed here.  That could be unreasonable because in addition to similarity the quality of a paper should be considered. However, we believe that the new metrics could be valuable in complementing the assessment of the impact of a given piece of work by reflecting in a comprehensive way its relationship with the literature. Furthermore, authors should be aware that some good practices of literature survey are not being adopted. 

\section{Conclusion and further work}

In an analysis of a considerably large corpus of 1,400 articles from two areas of knowledge, we have found that two important procedures in a literature survey are not followed by authors. It seems that other factors are important, which may not consider the scientific merit. This would be similar to the conclusion in~\cite{lea}, in which the number of citations and its impact may depend on the country of the authors or in~\cite{doestheplace} where the place of publication also had an effect on the citations. As for the method suggested here to evaluate the contributions from scientists, referred to as virtual scientometry, any type of bias is reduced. It is also in line with recent proposals that are based on a set of metrics to capture the importance of conferences~\cite{nivio}.
We also advocate that software tools should be developed to assist authors in performing a systematic search of the literature. In fact, the communities of medical doctors~\cite{emed} and software engineers~\cite{esoft} have now well-established methods to conduct surveys (problem formulation, studies selection and data collection are some examples), which are similar to what we propose here - albeit with distinct motivations.

\section{Acknowledgments}
The authors are grateful to FAPESP and CNPq (Brazil) for the financial support.

\end{document}